\documentclass[12pt,onecolumn,journal]{IEEEtran}

\usepackage{mathrsfs}
\usepackage{amsmath}
\usepackage{txfonts}
\usepackage[dvips]{graphicx}
\usepackage{graphicx}
\usepackage{caption}
\usepackage{subcaption}
\usepackage[ruled]{algorithm2e}

\ifCLASSINFOpdf

\else

\fi



\begin{document}

\title{Robust Compressed Sensing Under Matrix Uncertainties}

\author{Yipeng~Liu

\thanks{ Yipeng Liu is supported by the Fundamental Research Fund for the Central Chinese Universities (No. ZYGX2015KYQD004).
}
\thanks{Yipeng Liu is with School of Electronic Engineering / Center for Robotics
 / Center for Information in BioMedicine, University of Electronic Science and Technology of China (UESTC). Xiyuan Avenue 2006, Western High-Tech Zone, Chengdu, 611731, China. email: yipengliu@uestc.edu.cn  }%
\thanks{Manuscript received Month Day, 2015; revised Month Day, Year.}}

\markboth{Journal,~Vol.~X, No.~X, Month~Year}%
{Shell \MakeLowercase{\textit{et al.}}: Bare Demo of IEEEtran.cls for Journals}

\maketitle

\begin{abstract}
Compressed sensing (CS) shows that a signal having a sparse or compressible representation can be recovered from a small set of linear measurements. In classical CS theory, the sampling matrix and representation matrix are assumed to be known exactly in advance. However, uncertainties exist due to sampling distortion, finite grids of the parameter space of dictionary, etc. In this paper, we take a generalized sparse signal model, which simultaneously considers the sampling and representation matrix uncertainties. Based on the new signal model, a new optimization model for robust sparse signal reconstruction is proposed. This optimization model can be deduced with stochastic robust approximation analysis. Both convex relaxation and greedy algorithms are used to solve the optimization problem. For the convex relaxation method, a sufficient condition for recovery by convex relaxation is given; For the greedy algorithm, it is realized by the introduction of a pre-processing of the sensing matrix and the measurements. In numerical experiments, both simulated data and real-life ECG data based results show that the proposed method has a better performance than the current methods.

\end{abstract}


%

\IEEEpeerreviewmaketitle

\section{Introduction}
\label{sec1}



 Classical compressed sensing (CS) theory assumes that the representation matrix (dictionary) and sampling (measurement) matrix are known exactly in advance \cite{eldar_CS} \cite{candes_rup} \cite{donoho_cs}. However, some uncertainty or possible inaccuracy can affect them in many applications. For example, in the sparse representation of the signal, the assumed basis typically corresponds to a gridding of the parameter space, e. g., a discrete Fourier transformation (DFT) grid \cite{oppenheim_dtsp}. But in reality no physical field is exactly sparse in the DFT basis. No matter how finely the parameter space is gridded, the signal may not lie perfectly on the sampling points. This leads to mismatch between the assumed and the actual bases, which results in the uncertainty in the representation matrix. The sampling of the analogue signal's circuit noise and other non-linear effects can induce uncertainty in the sampling matrices \cite{liu_asd_cwss}. The classical sparse signal model did not consider these uncertainties; and the corresponding sparse signal recovery methods can suffer performance degeneration because of signal model mismatch.

Some papers have addressed related problems recently. \cite{herman_deviant}, \cite{herman2010mixed}, \cite{chi_basis_mismatch1} and \cite{chae_basis_mismatch2} analyzed signal recovery by basis pursuit (BP) and greedy algorithms with perturbations in either measurement matrix or representation matrix. To deal with the performance degeneration, several methods were proposed \cite{peyre_best_basis}-\cite{zhu_sparse_tls}. Instead of a fixed basis, \cite{peyre_best_basis} used a tree-structured dictionary of bases and the best bases were estimated with an iteratively processed recovery of the signal. \cite{liu_asd_cwss} and  \cite{rosenbaum2010sparse} relaxed the distortionless constraint to allow entry-wise sampling error by a series of large inequality zoom operations. Similarly, \cite{parker2011compressive} and \cite{krzakala2013compressed} generalized the approximate message passing (AMP) algorithm to hold the sampling matrix uncertainty with several parameters to be tuned or predefined.  \cite{yang_structured_perturbation} proposed a way only for the structured sensing matrix perturbation.  In \cite{stadler_missing_values} \cite{loh_missing_data}, two non-convex methods were proposed to deal with uncertainty in data in the sparse linear regression problem. The non-convexity requires knowledge of the $ \ell_1 $ norm of the unknown sparse signal in order to maintain bounded iterates, which is not available in many applications. \cite{zhu_sparse_tls} introduced a sparsely regularized total least-squares (SRTLS) method to deal with the uncertainty in the representation matrix. But its solver needs a number of iterations between the sparse signal estimation and the uncertainty matrix estimation, which implies a large computational burden. In summary, previous publications have not fully analyzed the resulting total uncertainty from both sampling and representation uncertainties. Furthermore, no algorithm of low computational complexity exists for sparse signal recovery in the presence of either sampling uncertainty or representation uncertainty.

In this paper, we generalize the sparse signal model containing both measurement and representation errors. Based on the generalized sparse signal model and possible statistical prior knowledge about the measurement and representation errors, a new data fitting constraint is deduced with stochastic uncertainty. We combine it with the $\ell_0 $ norm minimization based sparsity-inducing constraint, and obtain an optimization model for robust sparse signal recovery. Two approaches are used to solve the optimization problem. One relaxes the $\ell_0  $ norm to the $\ell_1 $ norm to obtain a convex programming problem; and the other one takes a greedy algorithm approach. For convex programming, we give a sufficient condition for successful recovery; and for the greedy algorithm, we prove it can be solved by regular greedy algorithms with transformations on sensing matrix and measurements. Numerical results show the performance of the proposed method with both simulated data and real-life ECG signals.

The rest of the paper is organized as follows. Section \ref{sec2} gives the generalized sparse signal model. In section \ref{sec3}, the corresponding optimization model for robust sparse signal recovery is deduced. In Section \ref{sec4}, both convex relaxation and a greedy algorithm are used to solve the optimization model. Section \ref{sec5} demonstrates the performance of the proposed method by numerical experiments. Finally, section \ref{sec6} presents the conclusions of this work.

\section{Generalized Sparse Signal Model}
\label{sec2}

In CS, instead of acquiring the signal ${\bf{x}} \in {\mathbb{R}^{N \times 1}} $ directly according to the Nyquist sampling, a measurement matrix ${\bf{\Phi }} \in {\mathbb{R}^{M \times N}}$ is used to sample the signal with $M \ll N$, which can be formulated as:
\begin{equation}
\label{eq2 measurement model}
{\bf{y}} = {\bf{\Phi x}},\
\end{equation}
where the obtained vector $ {\bf{y}} \in {\mathbb{R}^{M \times 1}}$  contains the sub-Nyquist-sampled random measurements.


Sparsity widely exists in many natural and man-made signals. It means that many of the representative coefficients are close to or equal to zero, when the signal is represented in a dictionary $ {\bf{\Psi }} \in {\mathbb{R}^{N \times N}}\ $. It can be formulated as:

\begin{equation}
\label{eq2 sparse model}
{\bf{x}} = {\bf{\Psi \theta }},\
\end{equation}
where $ {\bf{\theta }} \in {\mathbb{R}^{N \times 1}}\ $ is the representative vector with most of its entries are zero. When most of the entries are not strictly zero but trivial, or only a few of the entries are significant, strictly we should call the vector is compressible, but sometimes we say it be sparse too. The number of nonzero or significant entries are \emph{K}.

Combining (\ref{eq2 measurement model}) and (\ref{eq2 sparse model}), we can get:

\begin{equation}
\label{eq2 cs model}
{\bf{y}} = {\bf{\Phi \Psi \theta }} = {\bf{A\theta }},\
\end{equation}
where
\begin{equation}
\label{eq2 A model}
{\bf{A}} = {\bf{\Phi \Psi }},\
\end{equation}
 where \textbf{A} is called sensing matrix. Based on the standard sparse signal model (\ref{eq2 cs model}), classical CS proves that the signal can be successfully recovered by sparse signal recovery methods \cite{eldar_CS}.

To further consider the errors in the data, an additive noise term is included into the signal model as:
\begin{equation}
\label{eq2 cs model with additive noise}
{\bf{y}} = {\bf{A\theta }} + {\bf{n}},\
\end{equation}
where $ {\bf{n}} \in {\mathbb{R}^{M \times 1}} $ is the additive white Gaussian noise (AWGN) with zero mean and covariance matrix $ {\sigma ^2}{\bf{I}} $ \cite{kim2008gaussianity}.

However, in many practical scenarios, uncertainty in the sampling matrix exists. When sampling the analogue signals, uncertainty can result from various types of non-ideal effects, such as aliasing, aperture effect, jitter and deviation from the precise sample timing intervals, noise, and other non-linear effects. After sampling, uncertainty can also be introduced by an inconsistent channel effect, channels' coupling effect, and so on. Here we can model the sampling matrix with uncertainty as:
\begin{equation}
\label{eq2 sampling uncertainty}
{\bf{\Phi }} = {\bf{\bar \Phi }} + {{\bf{E}}_1},\
\end{equation}
where $ {\bf{\bar \Phi }} $ is the uncertainty-free sampling matrix which is known in advance or can be estimated by training data, and $ {{\bf{E}}_1} $  is the sampling matrix error. The exact information about $ {{\bf{E}}_1} $ cannot be available. We can approximately treat it as a random Gaussian variable matrix or some deterministic unknown variable matrix \cite{oppenheim_dtsp} \cite{kim2008gaussianity}.

There is uncertainty in the representation matrix (dictionary) too. It can result from the quantification of the representation matrix, such as the gridding of the parameter space of dictionary, the mismatch between the assumed dictionary for sparsity and the actual dictionary in which the signal is sparse, and so on. Similarly we model the representation matrix with uncertainty as:
\begin{equation}
\label{eq2 representaion uncertainty}
{\bf{\Psi }} = {\bf{\bar \Psi }} + {{\bf{E}}_2},\
\end{equation}
where $ {\bf{\bar \Psi }} $ is the uncertainty-free representation matrix which is known in advance or can be estimated by training data, and $ {{\bf{E}}_2} $ is the representation matrix error. We can approximately treat it as a random Gaussian variable matrix, or a random variable matrix in uniform distribution or some deterministic unknown variable matrix \cite{li_rab}.

To take the errors in both sampling and representation into consideration, we can reformulate (\ref{eq2 A model}) as:

\begin{equation}
\label{eq2 generalized A model}
\begin{array}{c}
{\bf{A}} = \left( {{\bf{\bar \Phi }} + {{\bf{E}}_1}} \right)\left( {{\bf{\bar \Psi }} + {{\bf{E}}_2}} \right)
 = \overline {\bf{A}}  + {\bf{E}},
\end{array}\
\end{equation}
where

\begin{equation}
\label{eq2 generalized A model 1}
\overline {\bf{A}}  = {\bf{\bar \Phi \bar \Psi }},\
\end{equation}

\begin{equation}
\label{eq2 generalized A model 2}
{\bf{E}} = {\bf{\bar \Phi }}{{\bf{E}}_2} + {{\bf{E}}_1}{\bf{\bar \Psi }} + {{\bf{E}}_1}{{\bf{E}}_2}.\
\end{equation}
\textbf{E} is the sensing matrix error. As can be seen in (\ref{eq2 generalized A model 2}), the correlation between measurement error $ {{\bf{E}}_1} $ and representation error $ {{\bf{E}}_2} $ affects the estimation of \textbf{E} mainly by the term $ {{\bf{E}}_1}{{\bf{E}}_2} $.

%
%

Based on the discussed model above, we can set up the sparse signal model with sampling and representation uncertainties and the additive noise. The generalized sparse signal model can be formulated as:

\begin{equation}
\label{eq2 generalized cs model}
\begin{array}{l}
{\bf{y}} = {\bf{A\theta }} + {\bf{n}},~~
{\bf{A}} = \overline {\bf{A}}  + {\bf{E}}.
\end{array}\
\end{equation}

\section{Optimization Model for Robust Sparse Signal Recovery}
\label{sec3}

\subsection{Classical methods}
\label{sec3.1}
Given the measurement vector \textbf{y} and the matrix $ {\bf{A}} $, we need to recover the sparse representative vector $ {\bf{\theta }} $. In CS, to find the sparsest signal that yields the measurements, we can solve the sparse least squares problem:

\begin{equation}
\label{eq3 L0 opt}
\begin{array}{c}
\mathop {\min }\limits_{\bf{\theta }} {\left\| {\bf{\theta }} \right\|_0},~~
{\rm{s}}{\rm{.~ t}}{\rm{.~~ }}{\bf{y}} = {\bf{A\theta }},
\end{array}
\end{equation}
where $ {\left\| {\bf{\theta }} \right\|_0} $ is the $ \ell_0 $ norm which counts the number of the nonzero entries of the vector $ {\bf{\theta }} $, and it encourages sparse distribution in $ \bf{\theta } $. It should be noted that the $ \ell_0 $ norm is not a full-fledged norm. Solving (\ref{eq3 L0 opt}) is NP-hard.

The basis pursuit denoising (BPDN) uses the $ \ell_1 $ norm to replace the $ \ell_0 $ norm to make it convex and relaxes the data fitting constraint to deal with the additive noise, where the the $ \ell_1 $ norm of the vector $ {\bf{\theta }} $ is defined as $ {\left\| {\bf{\theta }} \right\|_1} = \sum\nolimits_{n = 1}^N {\left| {{\theta _n}} \right|} $. BPDN can recover compressible signal with additive noise. However, it cannot allow the multiplicative error as in (\ref{eq2 generalized A model}) which is caused by sampling and representation uncertainties. In fact, the relaxed data fitting constraint of BPDN matches the sparse signal model with additive noise but does not match the generalized sparse signal model with sampling and representation uncertainties (\ref{eq2 generalized cs model}). The performance degradation of BPDN has been investigated in \cite{herman_deviant} \cite{chi_basis_mismatch1} \cite{chae_basis_mismatch2}.

\begin{figure}[htbp]
        \centering
                \begin{subfigure}[b]{0.25\textwidth}
                \includegraphics[width=\textwidth]{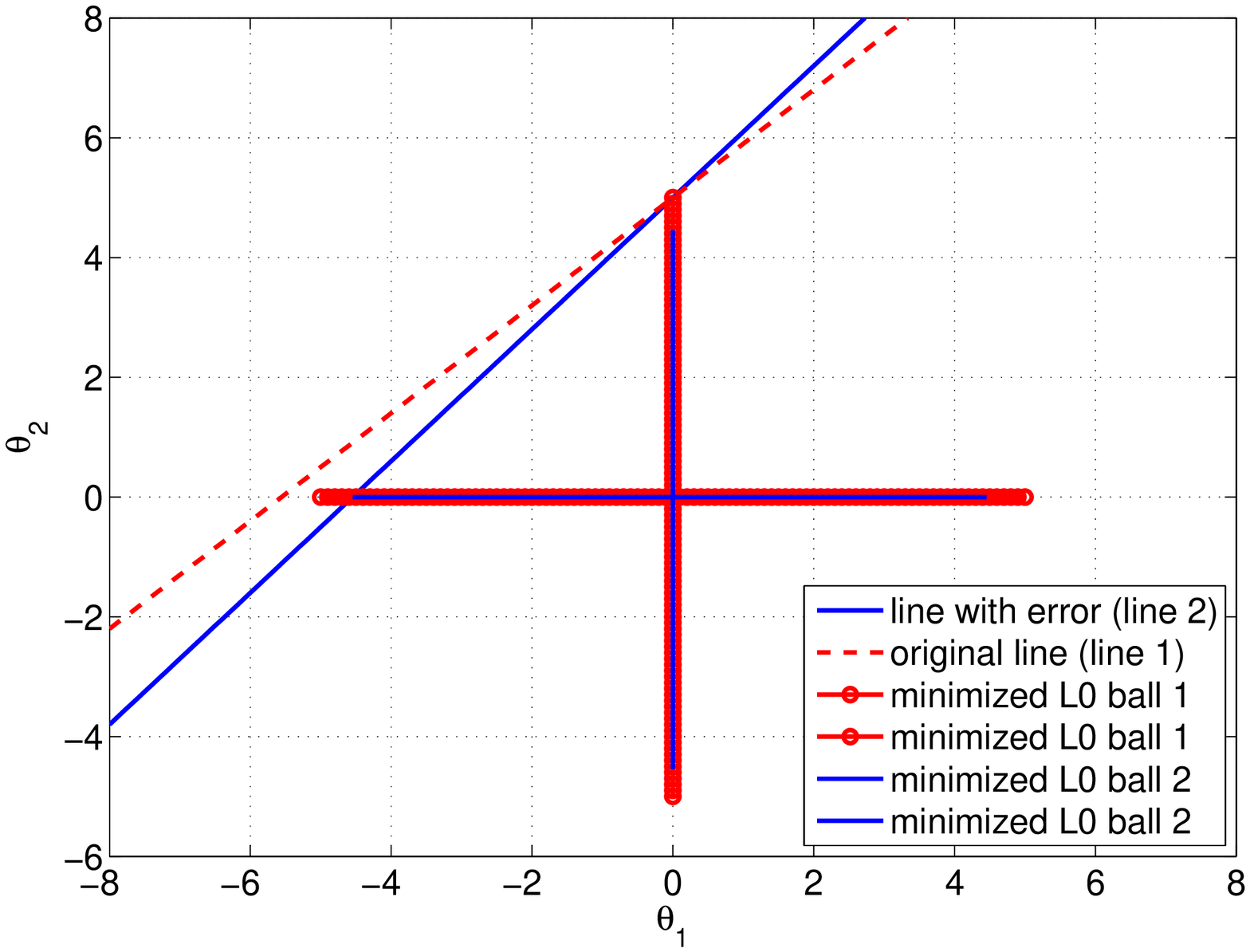}
                \caption{}
                \label{fig:L0 ball}
        \end{subfigure}%
        \begin{subfigure}[b]{0.25\textwidth}
                \includegraphics[width=\textwidth]{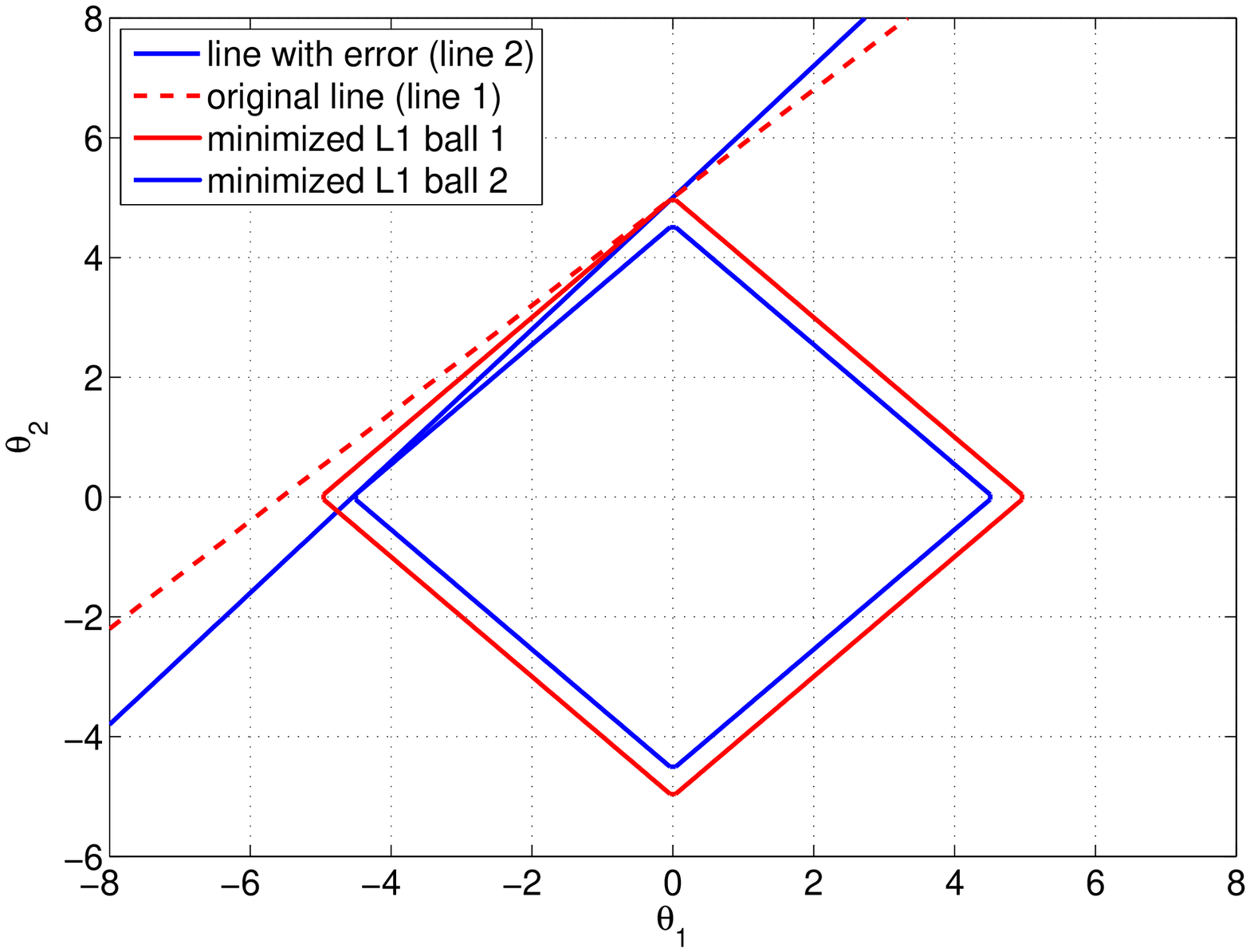}
                \caption{}
                \label{fig:L1 ball}
        \end{subfigure}%
        \caption{(a). The contour map of minimized $ \ell_0 $ balls which are tangent to the accurate line and the line which has error on slope (multiplicative noise): correct solution (red point of intersection ): $ {\left[ {0,~5} \right]^T} $; incorrect solution (blue point of intersection): $ {\left[ { -4.1667,~0} \right]^T} $; (b). The contour map of minimized $ \ell_1 $ balls which are tangent to the accurate line and the line which has error on slope (multiplicative noise): correct solution (red point of intersection): $ {\left[ {0,~5} \right]^T} $, incorrect solution (blue point of intersection): $ {\left[ { -4.1667,~0} \right]^T} $.}\label{fig:contour1}
\end{figure}

To explain why classical $ \ell_0  $ pseudo norm and $ \ell_1  $ norm based optimization methods could lead to incorrect solution with large error, we give an example to illustrate the situation in the presence of sampling and dictionary uncertainty, as shown in Fig. \ref{fig:contour1}. The designed data fitting constraint $ {\bf{y}} = {\bf{A\theta }} $ is $ {\theta _2} = 0.9{\theta _1} + 5 $ (i.e. $ 5 = \left[ {\begin{array}{*{20}{c}}
{ - 0.9}&1
\end{array}} \right]\left[ {\begin{array}{*{20}{c}}
{{\theta _1}}\\
{{\theta _2}}
\end{array}} \right] $) where $ {\bf{\theta }} = {\left[ {{\theta _1},{\theta _2}} \right]^T} $.  Because of multiplicative noise, the real data constraint in practice is $ {\theta _2} = 1.2{\theta _1} + 5 $ (i.e. $ 5 = \left[ {\begin{array}{*{20}{c}}
{ - 1.2}&1
\end{array}} \right]\left[ {\begin{array}{*{20}{c}}
{{\theta _1}}\\
{{\theta _2}}
\end{array}} \right]$). In Fig. \ref{fig:L0 ball} and Fig. \ref{fig:L1 ball}, we can see that the tangent points of the minimized $ \ell_0 $ and $ \ell_1 $ balls with the observed line are on the coordinate axes, which means the corresponding solutions are sparse. But they are far away from the ones of the minimized $ \ell_0 $ and $ \ell_1 $ balls with the original line which are the true solutions, which means that the error of the solutions are very large and they are not robust.

%

%
%
%

\subsection{Robust sparse optimization}

To robustly recover this generalized sparse signal, a new data fitting constraint, other than the one $ {\left\| {{\bf{y}} - \overline {\bf{A}} {\bf{\theta }}} \right\|_2^2} \le \varepsilon $ of BPDN, should be deduced. We assume the uncertainty term  \textbf{E} in (\ref{eq2 generalized A model}) is a random variable matrix, and \textbf{P} is the covariance matrix $ {\bf{P}} = {\rm E}\left( {{\bf{E}}^T{{\bf{E}}}} \right) $, which is positive-semidefinite and symmetric. We refer to the stochastic robust approximation \cite{ben_robust_optimization}. $ {{\bf{y}}_0} \in \mathbb{R} {^{M \times 1}} $ denotes the assumed measurement vector obtained by the signal model (\ref{eq2 generalized cs model}), i. e. $ {{\bf{y}}_0} = \left( {\overline {\bf{A}}  + {\bf{E}}} \right){\bf{\theta }} + {\bf{n}} $, and part of its parameters ($ {\overline {\bf{A}} } $ and \textbf{P}) are known in advance; and $ {\bf{y}} \in \mathbb{R} {^{M \times 1}} $ denotes the measurement vector obtained in practice without any knowledge of the signal model. We use the signal model to fit the practical measurements \emph{\textbf{y}}. We try to fit the obtained measurement vector with the generalized sparse signal model (\ref{eq2 generalized cs model}), and the expected value of the data fitting error can be formulated as:
\begin{equation}
\label{eq3.1 expected fitting error 1}
{\rm E}[\left\| {{\bf{y}} - {{\bf{y}}_0}} \right\|_2^2].\
\end{equation}
Incorporating the generalized sparse signal model (\ref{eq2 generalized A model}), we can get
\begin{equation}
\label{eq3.1 expected fitting error 2}
\begin{array}{c}
 {\rm E}\left[ {\left\| {{\bf{y}} - {{\bf{y}}_0}} \right\|_2^2} \right]
  = {\rm E}\left[ {{{\left( {{\bf{y}} - {{\bf{y}}_0}} \right)}^T}\left( {{\bf{y}} - {{\bf{y}}_0}} \right)} \right] \\
  = {\rm E}\left[ {{{\left[ {\left( {\overline {\bf{A}}  + {\bf{E}}} \right){\bf{\theta }} + {\bf{n}} - {\bf{y}}} \right]}^T}\left[ {\left( {\overline {\bf{A}}  + {\bf{E}}} \right){\bf{\theta }} + {\bf{n}} - {\bf{y}}} \right]} \right] \\
  = {\rm E}\left[ {\left\| {\overline {\bf{A}} {\bf{\theta }} - {\bf{y}}} \right\|_2^2 + {{\left( {\overline {\bf{A}} {\bf{\theta }} - {\bf{y}}} \right)}^T}\left( {{\bf{E\theta }} + {\bf{n}}} \right)} \right. \\
 \left. { + {{\left( {{\bf{E\theta }} + {\bf{n}}} \right)}^T}\left( {\overline {\bf{A}} {\bf{\theta }} - {\bf{y}}} \right) + \left\| {{\bf{E\theta }} + {\bf{n}}} \right\|_2^2} \right]. \\
 \end{array}
\end{equation}

Here we assume that all the entries in \textbf{n} are i.i.d Gaussian with $ {\rm E}\left( {{{\bf{n}}^T}{\bf{n}}} \right) = M{\sigma ^2} $, and \textbf{n} is independent from \textbf{E}. Thus we can get:

\begin{equation}
\label{eq3.1 expected fitting error 3}
\begin{array}{c}
{\rm E}\left\| {{\bf{y}} - {{\bf{y}}_0}} \right\|_2^2
 = {\rm E}\left[ {\left\| {\overline {\bf{A}} {\bf{\theta }} - {\bf{y}}} \right\|_2^2 + \left\| {{\bf{E\theta }} + {\bf{n}}} \right\|_2^2} \right]\\
 = \left\| {\overline {\bf{A}} {\bf{\theta }} - {\bf{y}}} \right\|_2^2 + {{\bf{\theta }}^T}{\bf{P\theta }} + {M\sigma ^2}
\end{array}.\
\end{equation}

Bounding this data fitting error expectation with a parameter $ \eta $ would give a new constraint which matches the generalized sparse signal model (\ref{eq2 generalized A model}) as:

\begin{equation}
\label{eq3.1 distortionless constraint}
\left\| {\overline {\mathbf{A}} {\mathbf{\theta }} - {\mathbf{y}}} \right\|_2^2 + {{\mathbf{\theta }}^T}{\mathbf{P\theta }} \leqslant \eta.
\end{equation}

Combining (\ref{eq3.1 distortionless constraint}) with the $ \ell_0 $ norm minimization yields the optimization model for recovering a generalized sparse signal with sampling and representation matrix uncertainties:
\begin{equation}
\label{eq3.1 rL0}
\begin{gathered}
  \mathop {\min }\limits_{\mathbf{\theta }} {\left\| {\mathbf{\theta }} \right\|_0},~~
  {\text{s}}{\text{.~t}}{\text{.  }}  \left\|  {\overline {\mathbf{A}} {\mathbf{\theta }} - {\mathbf{y}}} \right\|_2^2 + {{\mathbf{\theta }}^T}{\mathbf{P\theta }} \leqslant \eta.  \\
\end{gathered}
\end{equation}

It can be further generalized to:

\begin{equation}
\label{eq3 rl0 opt}
\mathop {\min }\limits_{\bf{\theta }} \left( {{\lambda _1}{{\left\| {\bf{\theta }} \right\|}_0} + {\lambda _2}{{\bf{\theta }}^T}{\bf{P\theta }} + \left\| {\overline {\bf{A}} {\bf{\theta }} - {\bf{y}}} \right\|_2^2} \right),
\end{equation}
where $ {\lambda _1} $ and $ {\lambda _2} $ are nonnegative parameters balancing the constraints, which can be tuned using cross validation, regularization path following etc. The proposed optimization (\ref{eq3 rl0 opt}) is called robust $ \ell_0 $ (RL0) optimization because it has robustness against the measurement and representation matrix uncertainties. One of its equivalent forms is:
\begin{equation}
\label{eq3 rl0 opt2}
\begin{gathered}
  \mathop {\min }\limits_{\mathbf{\theta }} {\left\| {{\mathbf{ \overline {\bf{A}} \theta }} - {\mathbf{y}}} \right\|_2} ,~~
  {\text{s}}{\text{.~t}}{\text{.~~}}{\left\| {\mathbf{\theta }} \right\|_0} \leqslant {\omega_1},{\text{ }}{{\mathbf{\theta }}^T}{\mathbf{P\theta }} \leqslant {\omega_2^2}, \\
\end{gathered}
\end{equation}
where $ \omega_1 $ and $ \omega _2 $ are parameters too.

We assume that the covariance matrix \textbf{P} is \emph{a priori} known in the RL0 optimization. On one hand, we can model \textbf{P} on the basis of an analysis of the CS setup. For example, we can make a possible assumption that the sampling matrix error is Gaussian as done in \cite{herman_deviant}; or, when the dictionary error is caused by finite gridding of the parameters (dictionary error corresponding to the quantization of the sparse vector) we can assume a uniform distribution of the gridding parameter, etc. On the other hand, we can estimate \textbf{P} as addressed in the errors-in-variables modeling literature, see e.g. \cite{fuller1987measurement}\cite{cheng1999statistical}\cite{van1991total}\cite{gallo1982consistency}. As far as we know, the best way to estimate P is via replicated observations \cite{fuller1987measurement}\cite{cheng1999statistical}\cite{van1991total}. Assuming that independent repeated measurements are available for each variable observed with error, this type of replications provides enough information about the error covariance matrix to derive a consistent unbiased estimate of \textbf{P}. A simple way to calculate \textbf{P} exists provided we can assume that the entries of the error matrix $ \left[ {\begin{array}{*{20}{c}}
{\bf{E}}&{\bf{n}}
\end{array}} \right] $  are i.i.d with mean zero and unknown variance $ {\delta ^2} $. The estimated covariance matrix is then $ {{\bf{P}}_{est}} = {\delta ^2}{\bf{I}} $. A consistent estimate of $ {\delta ^2} $  is provided by the squared minimal singular value  of $ \left[ {\begin{array}{*{20}{c}}
{{\bf{\bar A}}}&{\bf{y}}
\end{array}} \right] $  \cite{van1991total}. To address more general error distributions, we refer to the extended errors-in-variables modeling literature, see e.g.  \cite{cheng1999statistical} \cite{van1991total}. Using these consistent estimates instead of the true covariance matrix does not change the consistency properties of the parameter estimators for linear errors-in-variables models, of which the generalized sparse signal model (\ref{eq2 generalized cs model}) is a special case \cite{fuller1987measurement} \cite{gallo1982consistency}.

The newly proposed optimization model finds the sparsest solution among all possible solutions satisfying (\ref{eq3.1 distortionless constraint}). It can be formulated in a more generalized form as (\ref{eq3 rl0 opt}) which uses regularization parameters to balance different constraints.

\begin{figure}[htbp]
 \includegraphics[width=0.5\textwidth]{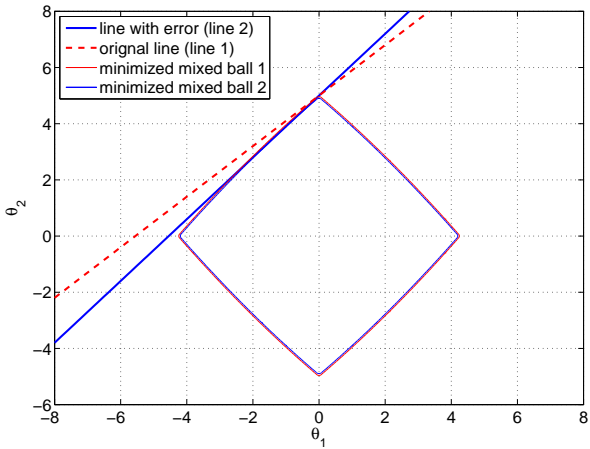}
 \caption{The contour of minimized mixed balls (L1 ball + ellipsoid) which are tangent to the accurate line and the line which has error on slope (multiplicative noise)): correct solution (red point of intersection): $ {\left[ {0,~5} \right]^T} $, incorrect solution (blue point of intersection): $ {\left[ { -0.5554,~4.3891} \right]^T} $.}
  \label{fig:mixed ball}
\end{figure}

If we further assume the random elements of \textbf{E} are uncorrelated, \textbf{P} can be a diagonal matrix. If we assume the entries in the multiplicative uncertainty matrix \textbf{E} are uncorrelated random variables with the variances $ {\delta _1},{\delta _2}, \cdots ,{\delta _N} $, (\ref{eq3 rl0 opt}) can be simplified as
\begin{equation}
\label{eq4.1 rl0 opt simplified}
\mathop {\min }\limits_{\bf{\theta }} {\lambda _1}{\left\| {\bf{\theta }} \right\|_0} + \left\| {{\bf{\bar A\theta }} - {\bf{y}}} \right\|_2^2 + {\lambda _2}\left\| {{\bf{\Delta \theta }}} \right\|_2^2,
\end{equation}
where
\begin{equation}
\label{eq4.1 rl0 opt simplified-2}
{\bf{\Delta }} = diag\left( {\begin{array}{*{20}{c}}
{\sqrt {{\delta _1}} }&{\sqrt {{\delta _2}} }& \cdots &{\sqrt {{\delta _N}} }
\end{array}} \right).
\end{equation}
When we further assume the multiplicative uncertainties are equal with the same variances, i.e. $ {\delta} = {\delta _1} = {\delta _2} =  \cdots  = {\delta _N} $, (\ref{eq4.1 rl0 opt simplified}) is further simplified in another form of the elastic net \cite{zou2005regularization}:
\begin{equation}
\label{eq4.1 elastic net}
\mathop {\min }\limits_{\bf{\theta }}  {{\lambda _1}{{\left\| {\bf{\theta }} \right\|}_0} + \left\| {\overline {\bf{A}} {\bf{\theta }} - {\bf{y}}} \right\|_2^2 + {\lambda _2}\delta \left\| {\bf{\theta}} \right\|_2^2}.
\end{equation}
Its performance for CS was evaluated in \cite{lai2013augmented} recently.

\section{Solutions}
\label{sec4}

Similarly to the classical sparse signal recovery methods, several kinds of methods can solve the optimization model (\ref{eq3 rl0 opt}). In this section, convex relaxation and a greedy algorithm are used to solve it.

\subsection{Convex relaxation}
\label{sec4.1}

A natural way relaxes the $ \ell_0 $ norm into the $ \ell_1 $ norm in (\ref{eq3 rl0 opt}), which achieves a convex optimization model:
\begin{equation}
\label{eq4.1 crl1 opt0}\mathop {\min }\limits_{\bf{\theta }} \left( {{\lambda _1}{{\left\| {\bf{\theta }} \right\|}_1} + {\lambda _2}{{\bf{\theta }}^T}{\bf{P\theta }} + \left\| {\overline {\bf{A}} {\bf{\theta }} - {\bf{y}}} \right\|_2^2} \right).
\end{equation}
This newly formed one is called convex robust $ \ell_1 $ (CR-L1) optimization. Another equivalent formulation, which is also the convex relaxation of (\ref{eq3 rl0 opt2}), is:
\begin{equation}
\label{eq4.1 crl1 opt}
\begin{gathered}
  \mathop {\min }\limits_{\mathbf{\theta }} {\left\| {{\mathbf{\overline {\bf{A}}\theta }} - {\mathbf{y}}} \right\|_2},~~
  {\text{s}}{\text{.~t}}{\text{.~~}}{\left\| {\mathbf{\theta }} \right\|_1} \leqslant {\omega_1},{\text{ }}{{\mathbf{\theta }}^T}{\mathbf{P\theta }} \leqslant {\omega_2^2}. \\
\end{gathered}
\end{equation}
Several approaches exist to solve the CR-L1 optimization, such as interior-point methods, subgradient methods, splitting Bregman algorithm, etc. The convergence can be guaranteed because of its convexity.

To explain why the proposed CR-L1 optimization (\ref{eq4.1 crl1 opt}) is robust to multiplicative noise, we use the same example as Fig. \ref{fig:contour1}. Thus the covariance matrix is $ {\bf{P}} = \left[ {\begin{array}{*{20}{c}}
{0.09}&0\\
0&0
\end{array}} \right] $. One example of the simplified CR-L1 optimization is:
\begin{equation}
\label{eq4.1 simplified crl1 opt}
\begin{array}{c}
\mathop {\min }\limits_{\bf{\theta }} \left( {{{\left\| {\bf{\theta }} \right\|}_1} + {{\bf{\theta }}^T}{\bf{P\theta }}} \right),~~
{\rm{s}}{\rm{.~ t}}{\rm{.~~  }}{\bf{y}} = {\bf{\overline {\bf{A}} \theta }}
\end{array}.
\end{equation}
 To combine the ellipsoid constraint's robustness to multiplicative noise and $ \ell_1 $ ball constraint's sparsity, we use the mixed ball in Fig. \ref{fig:mixed ball}. The tangent point of the minimized mixed ball ($ \ell_1 $ ball + ellipsoid) with the observed line is quite near the one of the minimized $ \ell_1 $ ball with the original line, and they are near the coordinate axes too. The additional quadratic term induces a slight compressibility loss but brings robustness to multiplicative noise in terms of better accuracy. Therefore, we can see that the proposed mixed ball can achieve a robust compressible solution.

For analysis' convenience, we assume there is no additive noise. Therefore, one equivalent form of (\ref{eq3 rl0 opt2}) is:
\begin{equation}
\label{eq4.1 simplified rl0 opt2}
\begin{gathered}
  \mathop {\min }\limits_{\mathbf{\theta }} {\left\| {\mathbf{\theta }} \right\|_0} + {\lambda _2}\sqrt {{{\mathbf{\theta }}^T}{\mathbf{P\theta }}} {\text{ }},~~
  {\text{s}}{\text{. t}}{\text{. }}{\mathbf{y}} = {\mathbf{ \overline {A} \theta }}. \\
\end{gathered}
\end{equation}

Similarly, one equivalent form of (\ref{eq4.1 crl1 opt}), which is also the convex relaxation of (\ref{eq4.1 simplified rl0 opt2}),  is:
\begin{equation}
\label{eq4.1 simplified crl1 opt2}
\begin{gathered}
  \mathop {\min }\limits_{\mathbf{\theta }} {\left\| {\mathbf{\theta }} \right\|_1} + {\lambda _2}\sqrt {{{\mathbf{\theta }}^T}{\mathbf{P\theta }}} {\text{ }},~~
  {\text{s}}{\text{. t}}{\text{. }}{\mathbf{y}} = {\mathbf{ \overline {A} \theta }}. \\
\end{gathered}
\end{equation}

\newtheorem{theorem}{Theorem}
\begin{theorem}[sufficient condition]\label{theorem: condition}
Assuming there is no additive noise, the CR-L1 optimization (\ref{eq4.1 simplified crl1 opt2}) can solve RL0 optimization (\ref{eq4.1 simplified rl0 opt2}) provided that
\begin{equation}
\label{eq4.1 condition_final}
M \geqslant \frac{{{{\left( {2\sqrt K  + \lambda_2 {C_2}} \right)}^2}}}{{C_1^2}}\log N ,
\end{equation}
where $ C_1 $ is a constant independent of the dimensions; $ C_2 = {\rm E}{\left\| {\mathbf{E}} \right\|_2} $ is the expectation of the compatible matrix norm of the $ \ell_2 $ vector norm.
\end{theorem}

\begin{IEEEproof}
Assuming the optimal solutions of the simplified RL0 optimization and CR-L1 optimization are:
\begin{equation}
\label{eq4.1 simplified rl0 opt:alpha}
{\mathbf{\alpha }} \in \mathop {\arg \min }\limits_{\mathbf{\theta }} {\left\| {\mathbf{\theta }} \right\|_0} + {\lambda _2}\sqrt {{{\mathbf{\theta }}^T}{\mathbf{P\theta }}} ,{\text{ s}}{\text{. t}}{\text{. }}{\mathbf{y}} = {\mathbf{ \overline {A} \theta }}
\end{equation}
and
\begin{equation}
\label{eq4.1 simplified crl1 opt:beta}
{\mathbf{\beta }} \in \mathop {\arg \min }\limits_{\mathbf{\theta }} {\left\| {\mathbf{\theta }} \right\|_1} + {\lambda _2}\sqrt {{{\mathbf{\theta }}^T}{\mathbf{P\theta }}} ,{\text{ s}}{\text{. t}}{\text{. }}{\mathbf{y}} = {\mathbf{ \overline {A} \theta }},
\end{equation}
the solutions of (\ref{eq4.1 simplified rl0 opt:alpha}) can solve (\ref{eq4.1 simplified crl1 opt:beta}), if
\begin{equation}
\label{eq4.1 condition}
{\left\| {{\mathbf{\alpha }} + {\mathbf{v}}} \right\|_1} + {\lambda _2}\sqrt {{{\left( {{\mathbf{\alpha }} + {\mathbf{v}}} \right)}^T}{\mathbf{P}}\left( {{\mathbf{\alpha }} + {\mathbf{v}}} \right)}  \geqslant {\left\| {\mathbf{\alpha }} \right\|_1} + {\lambda _2}\sqrt {{{\mathbf{\alpha }}^T}{\mathbf{P\alpha }}} ,\forall {\mathbf{v}} \in \ker ({\mathbf{\overline {A}}}),
\end{equation}
recalling
\begin{equation}
\label{eq4.1 kernel}
\ker \left( {\mathbf{\overline {A}}} \right) = \left\{ {{\mathbf{\theta }} \in {\mathbb{R}^N}:{\mathbf{\overline {A} \theta }} = 0} \right\}
\end{equation}
is the kernel (null space) of $ \mathbf{ \overline {A}} $. (\ref{eq4.1 condition}) means that in all the possible solutions of $ {\mathbf{y}} = {\mathbf{ \overline {A} \theta }} $, $ {\bf{\alpha }} $ also achieves the smallest value of $ {\left\| {\mathbf{\theta }} \right\|_1} + {\lambda _2}{{\mathbf{\theta }}^T}{\mathbf{P\theta }} $.

Let $ S $ be the support set $ S = \left\{ {n:{\alpha _n} \ne 0}, n = 1,2, \cdots , N \right\} $ and $ \overline S  = \left\{ {1, \cdots ,N} \right\}\backslash S $ where $ {\mathbf{\alpha }} = {\left[ {{\alpha _1},{\alpha _2}, \cdots ,{\alpha _N}} \right]^T} $, i.e. $ S $ is the support of the nonzero entries of $ {\mathbf{\alpha }} $; and $ \overline S $ is the support of the zero entries of $ {\mathbf{\alpha }} $. Then,
\begin{equation}
\label{eq4.1 condition: L1-1}
  \begin{gathered}
  {\left\| {{\mathbf{\alpha }} + {\mathbf{v}}} \right\|_1} = {\left\| {{{\mathbf{\alpha }}_S} + {{\mathbf{\alpha }}_{\overline S }} + {{\mathbf{v}}_S} + {{\mathbf{v}}_{\overline S }}} \right\|_1} \\
   = {\left\| {{{\mathbf{\alpha }}_S} + {{\mathbf{v}}_S}} \right\|_1} + {\left\| {{{\mathbf{v}}_{\overline S }}} \right\|_1} \\
   \geqslant {\left\| {{{\mathbf{\alpha }}_S}} \right\|_1} - {\left\| {{{\mathbf{v}}_S}} \right\|_1} + {\left\| {{{\mathbf{v}}_{\overline S }}} \right\|_1} \\
   = {\left\| {{{\mathbf{\alpha }}_S}} \right\|_1} + {\left\| {{{\mathbf{v}}_{\overline S }}} \right\|_1} + {\left\| {{{\mathbf{v}}_S}} \right\|_1} - 2{\left\| {{{\mathbf{v}}_S}} \right\|_1} \\
   = {\left\| {\mathbf{\alpha }} \right\|_1} + {\left\| {\mathbf{v}} \right\|_1} - 2{\left\| {{{\mathbf{v}}_S}} \right\|_1} \\
   \geqslant {\left\| {\mathbf{\alpha }} \right\|_1} + {\left\| {\mathbf{v}} \right\|_1} - 2\sqrt K {\left\| {\mathbf{v}} \right\|_2}, \\
\end{gathered}
\end{equation}
where $ {{\mathbf{v}}_S} $ keeps its entries corresponding to the support $ S $ and let the others be zero; and ${{{\bf{v}}_{\overline S }}}$ keeps its entries corresponding to the support $ {\overline S } $ and let the others be zero.

Furthermore, we have
\begin{equation}
\label{eq4.1 condition: ellipsoid-1}
\begin{gathered}
  \sqrt {{{\left( {{\mathbf{\alpha }} + {\mathbf{v}}} \right)}^T}{\mathbf{P}}\left( {{\mathbf{\alpha }} + {\mathbf{v}}} \right)}  = \sqrt {{{\left( {{\mathbf{\alpha }} + {\mathbf{v}}} \right)}^T}{\rm E}\left( {{\mathbf{E}}^T{{\mathbf{E}}}} \right)\left( {{\mathbf{\alpha }} + {\mathbf{v}}} \right)}  \\
   = {\rm E}{\left\| {{{\mathbf{E}}}\left( {{\mathbf{\alpha }} + {\mathbf{v}}} \right)} \right\|_2} \\
\end{gathered}
\end{equation}
and
\begin{equation}
\label{eq4.1 condition: ellipsoid-2}
\begin{gathered}
  {\rm E}{\left\| {{{\mathbf{E}}}\left( {{\mathbf{\alpha }} + {\mathbf{v}}} \right)} \right\|_2} = {\rm E}{\left\| {{{\mathbf{E}}}{\mathbf{\alpha }} + {{\mathbf{E}}}{\mathbf{v}}} \right\|_2} \\
   \geqslant {\rm E}{\left\| {{{\mathbf{E}}}{\mathbf{\alpha }}} \right\|_2} - {\rm E}{\left\| {{{\mathbf{E}}}{\mathbf{v}}} \right\|_2} \\
   \geqslant \sqrt {{{\mathbf{\alpha }}^T}{\mathbf{P\alpha }}}  - {\rm E}{\left\| {\mathbf{E}} \right\|_2}{\left\| {\mathbf{v}} \right\|_2} \\
   = \sqrt {{{\mathbf{\alpha }}^T}{\mathbf{P\alpha }}}  - C_2{\left\| {\mathbf{v}} \right\|_2}. \\
\end{gathered}
\end{equation}

Combining (\ref{eq4.1 condition: L1-1}) and (\ref{eq4.1 condition: ellipsoid-2}) results in:
\begin{equation}
\label{eq4.1 condition-all}
\begin{gathered}
  {\left\| {{\mathbf{\alpha }} + {\mathbf{v}}} \right\|_1} + {\lambda _2}\sqrt {{{\left( {{\mathbf{\alpha }} + {\mathbf{v}}} \right)}^T}{\mathbf{P}}\left( {{\mathbf{\alpha }} + {\mathbf{v}}} \right)}  \\
   \geqslant {\left\| {\mathbf{\alpha }} \right\|_1} + {\left\| {\mathbf{v}} \right\|_1} - 2\sqrt K {\left\| {\mathbf{v}} \right\|_2} + {\lambda _2}\sqrt {{{\mathbf{\alpha }}^T}{\mathbf{P\alpha }}}  - {\lambda _2}{C_2}{\left\| {\mathbf{v}} \right\|_2} \\
   = {\left\| {\mathbf{\alpha }} \right\|_1} + {\lambda _2}\sqrt {{{\mathbf{\alpha }}^T}{\mathbf{P\alpha }}}  + {\left\| {\mathbf{v}} \right\|_1} - \left( {2\sqrt K  + {\lambda _2}{C_2}} \right){\left\| {\mathbf{v}} \right\|_2}. \\
\end{gathered}
\end{equation}

From (\ref{eq4.1 condition-all}), we can see that (\ref{eq4.1 condition}) holds provided that $ {\left\| {\bf{v}} \right\|_1} \ge ( 2\sqrt K + \lambda_2 C_2 ) {\left\| {\bf{v}} \right\|_2} $. In general we have $ 1 \le {{{{\left\| {\bf{v}} \right\|}_1}} \mathord{\left/
 {\vphantom {{{{\left\| {\bf{v}} \right\|}_1}} {{{\left\| {\bf{v}} \right\|}_2}}}} \right.
 \kern-\nulldelimiterspace} {{{\left\| {\bf{v}} \right\|}_2}}} \le \sqrt N  $. However, if the elements of $ {\mathbf{\overline {A}}} \in {\mathbb{R}^{M \times N}} $ are sampled i.i.d. from Gaussian process with zero mean and unit variance, with high probability, we have
\begin{equation}
\label{eq4.1 condition: L1-2}
\frac{{{{\left\| {\mathbf{v}} \right\|}_1}}}{{{{\left\| {\mathbf{v}} \right\|}_2}}}  \le  \frac{{C_1\sqrt M }}{{\sqrt {\log \frac{N}{M}} }}{\text{, ~for~ all~ }}{\mathbf{v}} \in \ker \left( {\mathbf{\overline {A}}} \right),
\end{equation}
where $ C_1 $ is a constant \cite{kashin1977diameters} \cite{garnaev1984widths}. When $ {\mathbf{\overline {A}}} $ is Gaussian, with high probability, we have (\ref{eq4.1 condition}) holds if
\begin{equation}
\label{eq4.1 exact condition_final}
M \ge \left( {\log N - \log M} \right){\left( {\frac{{2\sqrt K  + {\lambda _2}{C_2}}}{{{C_1}}}} \right)^2}
\end{equation}
is satisfied.

Obviously we have $ M \ge 1 $. (\ref{eq4.1 exact condition_final}) can be met if (\ref{eq4.1 condition_final}) holds. Therefore, Theorem \ref{theorem: condition} is proved.
\end{IEEEproof}

The similar sufficient condition for convex relaxation of (\ref{eq4.1 elastic net}) can be obtained if we let $ C_2 = 1 $ in (\ref{eq4.1 condition_final}). Furthermore, if $ C_2 = 0 $ which means no noise in the model, the sufficient condition for standard BP for CS can be obtained, and the resulted condition agrees with previous conclusions too \cite{eldar_CS} \cite{candes_rup} \cite{donoho_cs}. In order to suppress the multiplicative noise, the proposed new method has a larger lower bound on the required number of measurements than that of the standard BP with no multiplicative noise. The requirement of additional measurements is the price paid for multiplicative noise suppression.

The introduction of the $ \ell_2 $ norm constraint will not only enhance robustness, but also smooth the $ \ell_1 $ norm based penalty function, as can be seen in Fig. \ref{fig:mixed ball} \cite{nesterov2005smooth}. The convergence of the sub-gradient algorithm would be accelerated.

\subsection{Greedy algorithm}
\label{sec4.2}

To reduce the computational complexity, greedy algorithms can be used to solve the RL0 optimization model. In contrast to the classical OMP which greedily chooses the atoms giving the minimum $ \left\| {{\mathbf{y}} - {\mathbf{\overline {A}\theta }}} \right\|_2^2 $ , we update by choosing the ones to minimize

\begin{equation}
\begin{gathered}
  f\left( {\mathbf{\theta }} \right) = \left\| {{\mathbf{y}} - {\mathbf{\overline {A}\theta }}} \right\|_2^2 + {{\mathbf{\theta }}^T}{\mathbf{P\theta }} \\
   = {\left( {{\mathbf{y}} - {\mathbf{\overline {A}\theta }}} \right)^T}\left( {{\mathbf{y}} - {\mathbf{\overline {A}\theta }}} \right) + {{\mathbf{\theta }}^T}{\mathbf{P\theta }} \\
   = {{\mathbf{y}}^T}{\mathbf{y}} - 2{{\mathbf{y}}^T}{\mathbf{\overline {A}\theta }} + {{\mathbf{\theta }}^T}\left( {{{\mathbf{\overline {A}}}^T}{\mathbf{\overline {A}}} + {\mathbf{P}}} \right){\mathbf{\theta }}. \\
\end{gathered}
	\label{eq:4.2:object function}
\end{equation}
To find the minimum with different values of $ \theta $, we can let
\begin{equation}
\begin{gathered}
  \frac{{\partial f\left( {\mathbf{\theta }} \right)}}{{\partial {\mathbf{\theta }}}} =  - 2{{\mathbf{y}}^T}{\mathbf{\overline {A}}} + {{\mathbf{\theta }}^T}\left[ {\left( {{{\mathbf{\overline {A}}}^T}{\mathbf{\overline {A}}} + {\mathbf{P}}} \right) + {{\left( {{{\mathbf{\overline {A}}}^T}{\mathbf{\overline {A}}} + {\mathbf{P}}} \right)}^T}} \right] \\
   =  - 2{{\mathbf{y}}^T}{\mathbf{\overline {A}}} + 2{{\mathbf{\theta }}^T}\left( {{{\mathbf{\overline {A}}}^T}{\mathbf{\overline {A}}} + {\mathbf{P}}} \right) \\
   = 0, \\
\end{gathered}
	\label{eq:4.2:differential operation}
\end{equation}
which results in a new equation:
\begin{equation}
{\mathbf{B\theta }} = {\mathbf{z}},
\label{eq:4.2:linear inverse}
\end{equation}
where
\begin{equation}
{\mathbf{B}} =  {{{\mathbf{\overline {A}}}^T}{\mathbf{\overline {A}}} + {\mathbf{P}}} \\= {\left( {{\mathbf{ \overline{\Phi} \overline{\Psi} }}} \right)^T}{\mathbf{\overline{\Phi} \overline{\Psi} }} + {\mathbf{P}},
\label{eq:4.2:B}
\end{equation}
\begin{equation}
{\mathbf{z}} = {{\mathbf{\overline {A}}}^T}{\mathbf{y}}.
\label{eq:4.2:z}
\end{equation}
Therefore, in greedy algorithms, we can find one or several atoms which give the minimum residual for each iteration. i.e. We use the new "sensing matrix" \textbf{B} and "measurements" \textbf{z} instead of $ \mathbf{\overline {A}} $ and \textbf{y}. With these transformations of sensing matrix and measurement, we can use all the greedy algorithms for CS as before \cite{tropp2004greed}. We should note that the new sensing matrix \textbf{B} should not be fully random but partly random. The component \textbf{P} is deterministic and may prevail when the multiplicative error is strong and its corresponding covariance matrix has a bad CS performance, such as a large coherence \cite{candes2008introduction}, a large restricted isometry constant (RIC) \cite{blanchard2011compressed}. It can require a stricter condition for successful recovery.

A generalized OMP (orthogonal matching pursuit), which is also called OMMP (orthogonal multi-matching pursuit), is used to realize the robust greedy algorithm \cite{xu2012performance} \cite{shim2012generalized}. It is in the sense that multiple indices are identified in each iteration. When the number of identified indices is $ \rho =1 $, OMMP is equivalent to OMP. The proposed robust orthogonal multiple matching pursuit (ROMMP) algorithm is summarized in Algorithm \ref{alg1}. The algorithm can be stopped when the residual is smaller than a threshold $ \epsilon $ which is proportional to the standard deviation of its additive Gaussian noise  \cite{cai2011orthogonal}.
\begin{algorithm}[htbp]
\small
\caption{robust orthogonal multiple matching pursuit}
\label{alg1}
$\bullet$ \textbf{Input:}  $ \bf{ \overline {\Psi}} $,  $\bf{B}$ in (\ref{eq:4.2:B}), $\bf{z}$ in (\ref{eq:4.2:z}) and $ \rho $

$\bullet$ \textbf{Output:} $\hat{\mathbf{x}}$

$\bullet$ Initial Iteration: $t := 0$

$\bullet$ Initial Support: $\hat{\Omega}_t = \emptyset$

$\bullet$ Initial Residual: $\mathbf{r}_t = \mathbf{z}$

\Repeat{$ ||\mathbf{r}_t||_2 \leq \epsilon $}
{
$\bullet$ Set $t := t + 1$;

$\bullet$ Update Support: $\hat{\Omega}_t = \hat{\Omega}_{t-1} \cup \lbrace \underset{\textbf{i} \notin \hat{\Omega}_{t-1}}{argmax} \: |(\textbf{B})_\textbf{i}^T \{\mathbf{r_t}\}|\rbrace$

$\bullet$ Update Coefficients: $\hat{\bf{\theta}_t} = \underset{\bf{\vartheta}}{argmin} ||\mathbf{z}-(\textbf{B})_{\Omega_t} \vartheta  ||_2$

$\bullet$ Calculate Residual: $\bf{r}_t = \mathbf{z} - \textbf{B} \bf{\theta}_t$
}

$\bullet$ $\hat{\mathbf{x}} = \bf{ \overline {\Psi}} \hat{\bf{\theta}_t}$

\end{algorithm}

%

\section{Numerical Experiments}
\label{sec5}


\subsection{Simulated data}
\label{sec5.1}
In the numerical experiments with simulated data, the length of the sparse signal $ {\bf{\theta }} $ is \emph{N} = 200. It contains only a few nonzero entries. The number of nonzero entries of the sparse signal \emph{K} = 10. The locations of the nonzero entries vary randomly. It is normalized by its $ \ell_2 $ norm. The signal is sparse with respect to the canonical basis of the Euclidean space, i. e. $ \overline {\bf{\Psi }}  = {{\bf{I}}_{N \times N}} $; and the sampling matrix $ \overline {\bf{\Phi }} $ is Gaussian distributed. The matrix $ {\bf{A}} $  are be generated by the signal model (\ref{eq2 generalized A model 2}) and (\ref{eq2 generalized cs model}), where $ \bar \Phi $ and $ \bf{E}_1 $ are generated by sampling a white Gaussian distribution with zero mean, $ \bar \Psi $ is an identity matrix, and $ \bf{E}_2 $ is generated by sampling a uniform distribution with zero mean. To make the expression of the signal-to-multiplicative-noise ratio in the signal model convenient, Every column of $ \overline{\bf{A}} $ and $ \bf{E} $ is normalized by its $ \ell_2 $ norm, but a uncertainty parameter $ \tau  \in \ \mathbb{R} $ is employed to weight the multiplicative noise matrix,  i.e. $ {\bf{A}} = \overline {\bf{A}}  + \tau \bf{E} $. The standard deviation of the AWGN \textbf{n} is $ \sigma  = 0.1 $.

All the parameters in all three methods are chosen to give the best performance based on advanced searching. Here $ \upsilon $, $ {\lambda} $, $ {\lambda _1} $ and $ {\lambda _2} $ are chosen to achieve the best accuracy performance. Getting these optimal values for the parameters is not straightforward. Similarly to parameter estimation in BPDN, cross validation may be used resulting in additional computation burden. The number of iterations for SRTLS is 20. Before the number of iterations reach 20, the error does not vary much. The matrix \textbf{P} is chosen as the sampled covariance matrix $ {\bf{P}} = {{\sum\nolimits_{l = 1}^L {{\bf{U}}_l^T{{\bf{U}}_l}} } \mathord{\left/
 {\vphantom {{\sum\nolimits_{l = 1}^L {{\bf{U}}_l^T{{\bf{U}}_l}} } L}} \right.
 \kern-\nulldelimiterspace} L} $, where \emph{L} is the number of Monte Carlo simulations, which is chosen to be \emph{L} = 500.

To quantify the performance of signal recovery, the estimation error is calculated via the normalized mean L-b error:
\begin{equation}
\label{eq4 Lq error}
{e_b} = \frac{1}{{2L}}\sum\limits_{l = 1}^L {\frac{{{{\left\| {{{\bf{x}}_l} - {{{\bf{\hat x}}}_l}} \right\|}_b}}}{{{{\left\| {{{\bf{x}}_l}} \right\|}_b}}}}
\end{equation}
and the mean coherence:
\begin{equation}
\label{eq4 coherence error}
c = \frac{1}{L}\sum\limits_{l = 1}^L {\frac{{\left| {{{\bf{x}}_l}^H{{{\bf{\hat x}}}_l}} \right|}}{{{{\left\| {{{\bf{x}}_l}} \right\|}_2}{{\left\| {{{{\bf{\hat x}}}_l}} \right\|}_2}}}},
\end{equation}
where $ {{\bf{x}}_l} $ and $ {{\bf{\hat x}}_l} $ are the real and estimated signals in the \emph{l}-th experiment, and they are normalized by their $ \ell_b $ norms; $ b \in \{ 1,2\} $ indicates different criteria for the evaluation of the estimation performance; when \emph{b} = 1, we call $ {e_1} $ the normalized mean L1 error; and when \emph{b} = 2, $ {e_2} $ is called the normalized mean L2 error.

\begin{figure}[htbp]
\centering
        \begin{subfigure}[b]{0.48\textwidth}
                \includegraphics[width=9cm,height=5cm]{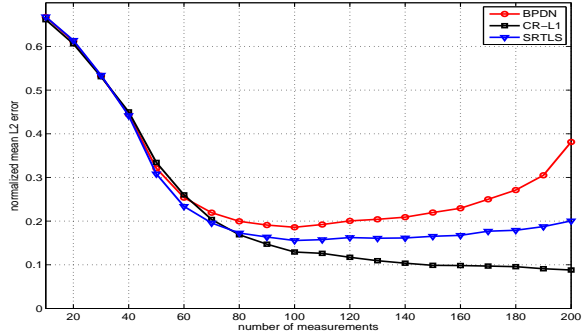}
                \caption{normalized mean L2 error vs number of measurements}
                \label{fig:1-1}
        \end{subfigure}%

        ~ 
        \begin{subfigure}[b]{0.48\textwidth}
                \includegraphics[width=9cm,height=5cm]{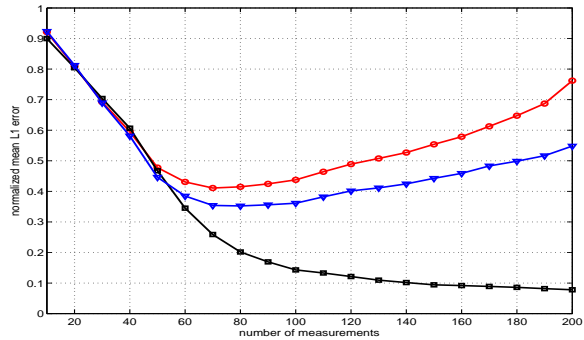}
                \caption{normalized mean L1 error vs number of measurements}
                \label{fig:1-2}
        \end{subfigure}

        ~ 
        \begin{subfigure}[b]{0.48\textwidth}
                \includegraphics[width=9cm,height=5cm]{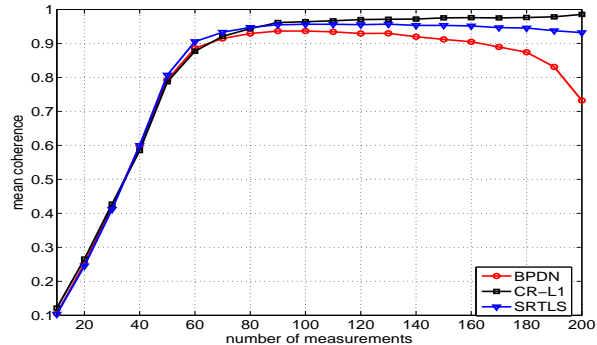}
                \caption{mean coherence vs number of measurements}
                \label{fig:1-3}
        \end{subfigure}
        \caption{Experiments of Group A: The normalized mean L1 and L2 errors and mean coherence versus the number of measurements when the uncertainty parameter $ \tau  = 0.3\ $.}
\label{figure1}
\end{figure}

\begin{figure}[htbp]
\centering
        \begin{subfigure}[b]{0.48\textwidth}
                \includegraphics[width=9cm,height=5cm]{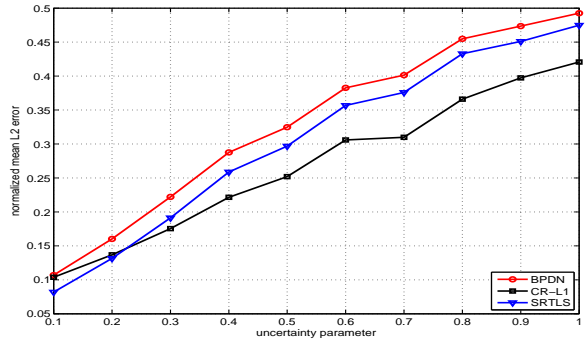}
                \caption{normalized mean L2 error vs uncertainty parameter}
                \label{fig:3-1}
        \end{subfigure}%

        ~ 
        \begin{subfigure}[b]{0.48\textwidth}
                \includegraphics[width=9cm,height=5cm]{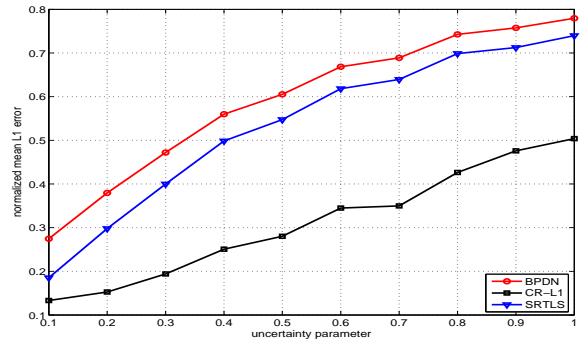}
                \caption{normalized mean L1 error vs uncertainty parameter}
                \label{fig:3-2}
        \end{subfigure}

        ~ 
        \begin{subfigure}[b]{0.48\textwidth}
                \includegraphics[width=9cm,height=5cm]{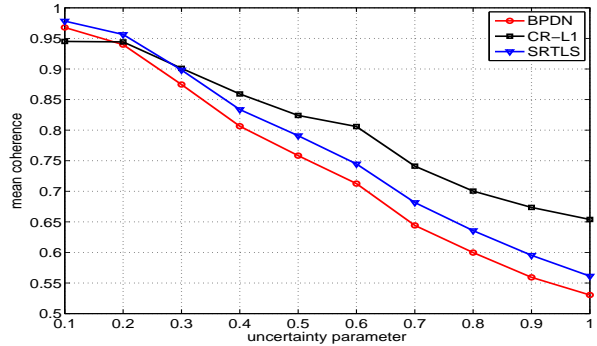}
                \caption{mean coherence vs uncertainty parameter}
                \label{fig:3-3}
        \end{subfigure}
        \caption{Experiments of Group A: The normalized mean L1 and L2 errors and mean coherence versus the uncertainty parameter when  the number of measurements \emph{M} = 100.}
\label{figure3}
\end{figure}

Fig. \ref{figure1} - Fig. \ref{figure3} demonstrate the signal reconstruction performance with the normalized mean L1 and L2 errors, and mean coherence. Fig. \ref{figure1} gives the normalized mean L1 and L2 errors and mean coherence with the number of measurements ranging from \emph{M} = 10 to 200, when the uncertainty parameter $ \tau  = 0.3 $; Fig. \ref{figure3} gives the normalized mean L1 and L2 errors and mean coherence with the uncertainty parameter ranging from $ \tau  = 0.1 $ to $ \tau  = 1 $, when the number of measurements \emph{M} = 100.

%
%
%
%
%
%

From Fig. \ref{figure1}, we can see that the CR-L1 optimization outperforms or at least share the similar performance with the BPDN and SRTLS with all the possible number of measurements in terms of the normalized mean L1 and L2 errors and mean coherence, especially when the number of measurements is large. Fig. \ref{figure3} also shows that the CR-L1 optimization performs the best or achieves similar performance, and the performance improvement is especially obvious when the uncertainties are strong.


\subsection{ECG data}
\label{sec5.2}

To test the proposed method for real-life data, we use ECG data which is obtained from the Physiobank database \cite{goldberger2000physiobank} \cite{mamaghanian2011compressed} \cite{liu2013multi}. Mobile ECG monitoring is one of the most popular applications in CS of ECG signals. In this application, the computational complexity should be as low as possible. Therefore, in this group of numerical experiments, the greedy algorithms, i.e. OMP, OMMP with $ \rho =4 $, and the proposed ROMMP with $ \rho =4 $, are compared. The measurement matrix is the Gaussian matrix. The utilized representation matrix is given by the orthogonal Daubechies wavelets (db 10) with the decomposition level 5 which is one of the most popular wavelet families for ECG compression \cite{liu2013multi}. The ECG data has 15 channels with 37888 samples for each channel. In each channel, the data are divided into 37 segments, i.e. the length of the signal in each reconstruction is \emph{N} = 1024. The sensing matrix error is generated similarly to that in section \ref{sec5.1}, but only the dictionary uncertainty matrix is set to be 0. In addition, the standard deviation of the AWGN \textbf{n} is $ \sigma  = 0.30 $.



\begin{figure}[!h]
 \centering
 \includegraphics[angle= 0, scale = 0.40]{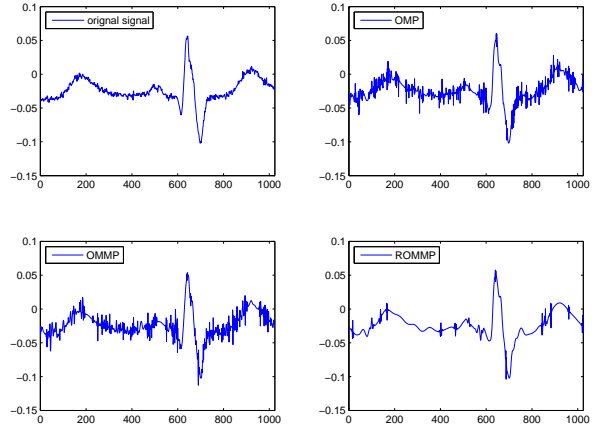}
 \caption{ECG signal estimates from the sub-samples.}
 \label{figecg}
\end{figure}

Fig.~\ref{figecg} shows part of an ECG signal and its estimates from sub-samples by OMP, OMMP and ROMMP when the number of the compressive measurements is \emph{M} = 410  and   $ {\tau  } = 0.1 $. We can see that the signal reconstructed by ROMMP is less noisy than the ones reconstructed by the OMP and OMMP.

\begin{figure}[htbp]
\centering
        \begin{subfigure}[b]{0.48\textwidth}
                \includegraphics[width=9cm,height=5cm]{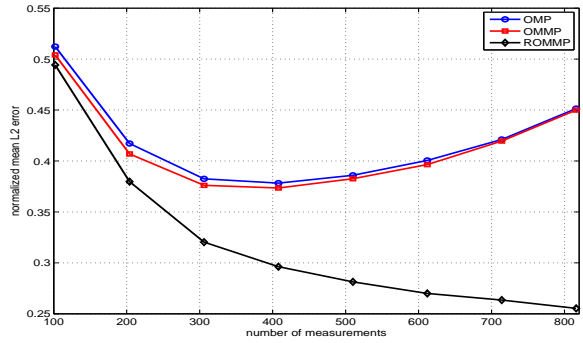}
                \caption{normalized mean L2 error vs number of measurements}
                \label{fig:20-1}
        \end{subfigure}%

        ~ 
        \begin{subfigure}[b]{0.48\textwidth}
                \includegraphics[width=9cm,height=5cm]{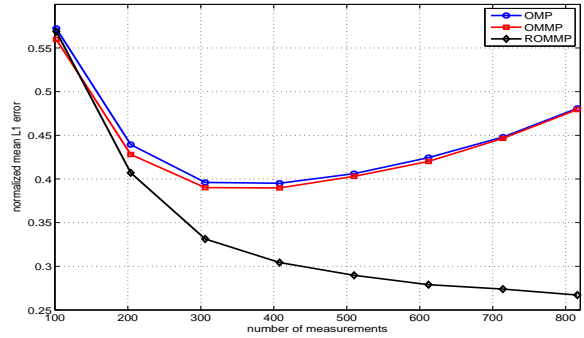}
                \caption{normalized mean L1 error vs number of measurements}
                \label{fig:20-2}
        \end{subfigure}

        ~ 
        \begin{subfigure}[b]{0.48\textwidth}
                \includegraphics[width=9cm,height=5cm]{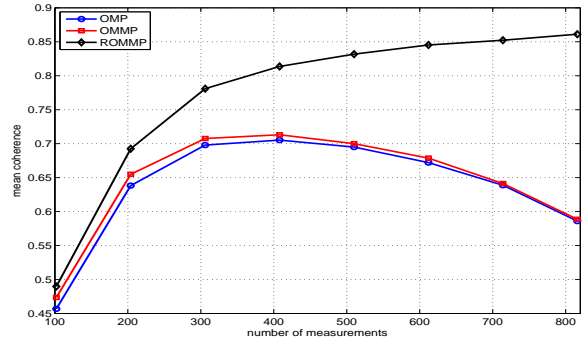}
                \caption{mean coherence vs number of measurements}
                \label{fig:20-3}
        \end{subfigure}
        \caption{Experiments of Group B: The normalized mean L1 and L2 errors and mean coherence values with various number of measurements for a fixed
uncertainty parameter.}\label{figure20}
\end{figure}

\begin{figure}[htbp]
\centering
\begin{subfigure}[b]{0.48\textwidth}
                \includegraphics[width=9cm,height=5cm]{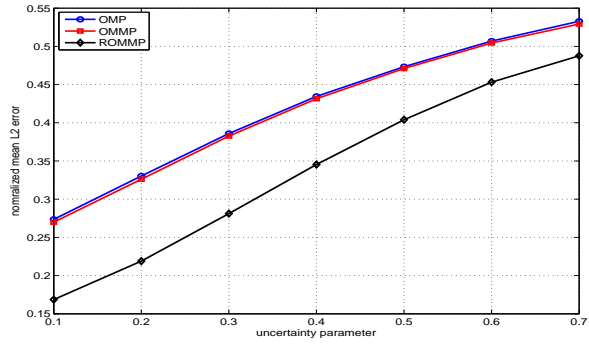}
                \caption{normalized mean L2 error vs vs uncertainty parameter}
                \label{fig:21-1}
        \end{subfigure}%

        ~ 
        \begin{subfigure}[b]{0.48\textwidth}
                \includegraphics[width=9cm,height=5cm]{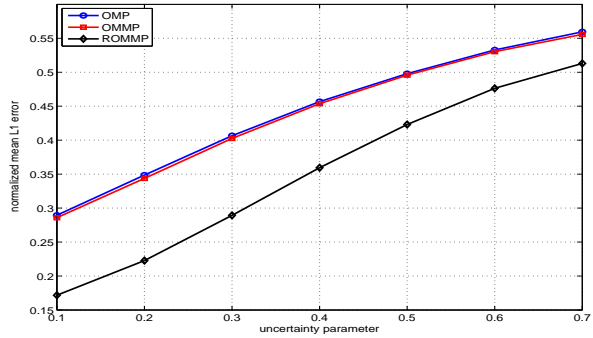}
                \caption{normalized mean L1 error vs uncertainty parameter}
                \label{fig:21-2}
        \end{subfigure}

        ~ 
        \begin{subfigure}[b]{0.48\textwidth}
                \includegraphics[width=9cm,height=5cm]{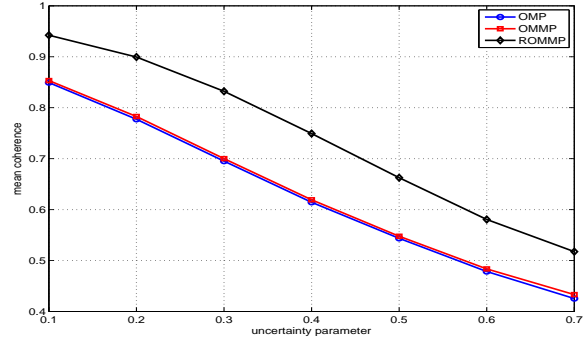}
                \caption{mean coherence vs uncertainty parameter}
                \label{fig:21-3}
        \end{subfigure}
\caption{Experiments of Group B: The normalized mean L1 and L2 errors and mean coherence with various uncertainty parameter for a fixed number of measurements.}
\label{figure21}
\end{figure}

In Fig.~\ref{figure20}, the normalized mean L1 and L2 errors and mean coherence for the reconstruction of ECG signal from the noisy compressive measurements with different number of measurements is shown when the uncertainty parameter is $ \tau  = 0.3 $; and Fig.~\ref{figure21} shows the normalized mean L1 and L2 errors and mean coherence for the reconstruction of the ECG signal from the noisy compressive measurements with different uncertainty parameters when the number of measurements are 512. It can be seen that OMP and OMMP have almost the same reconstruction accuracy. However, ROMMP improves the accuracy much better over various number of measurements and uncertainty degrees compared to OMP and OMMP.

\begin{figure}[htbp]
\centering
\begin{subfigure}[b]{0.48\textwidth}
                \includegraphics[width=9cm,height=5cm]{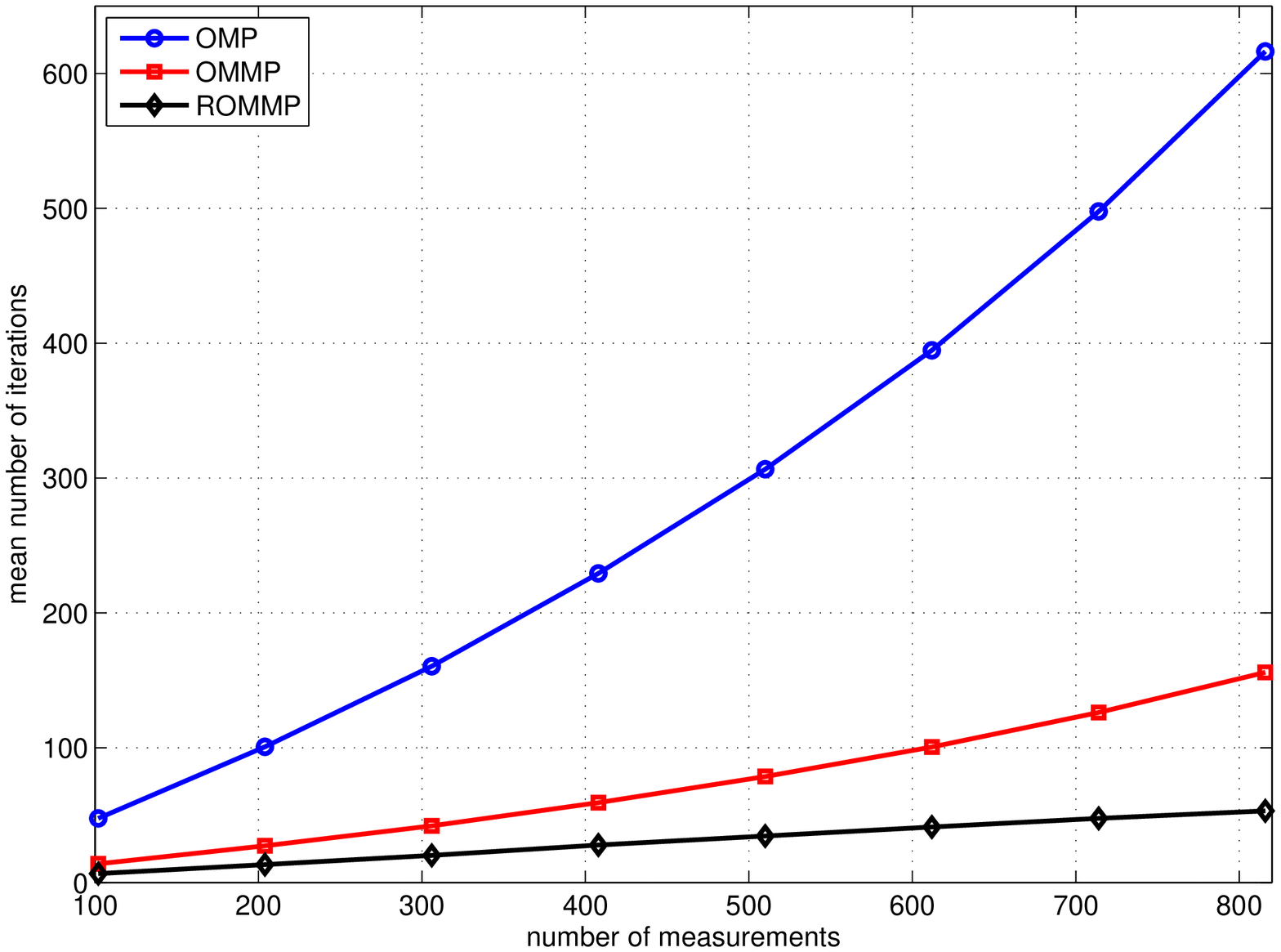}
                \caption{mean number of iterations vs number of measurements}
                \label{fig:22-1}
        \end{subfigure}%

        ~ 
        \begin{subfigure}[b]{0.48\textwidth}
                \includegraphics[width=9cm,height=5cm]{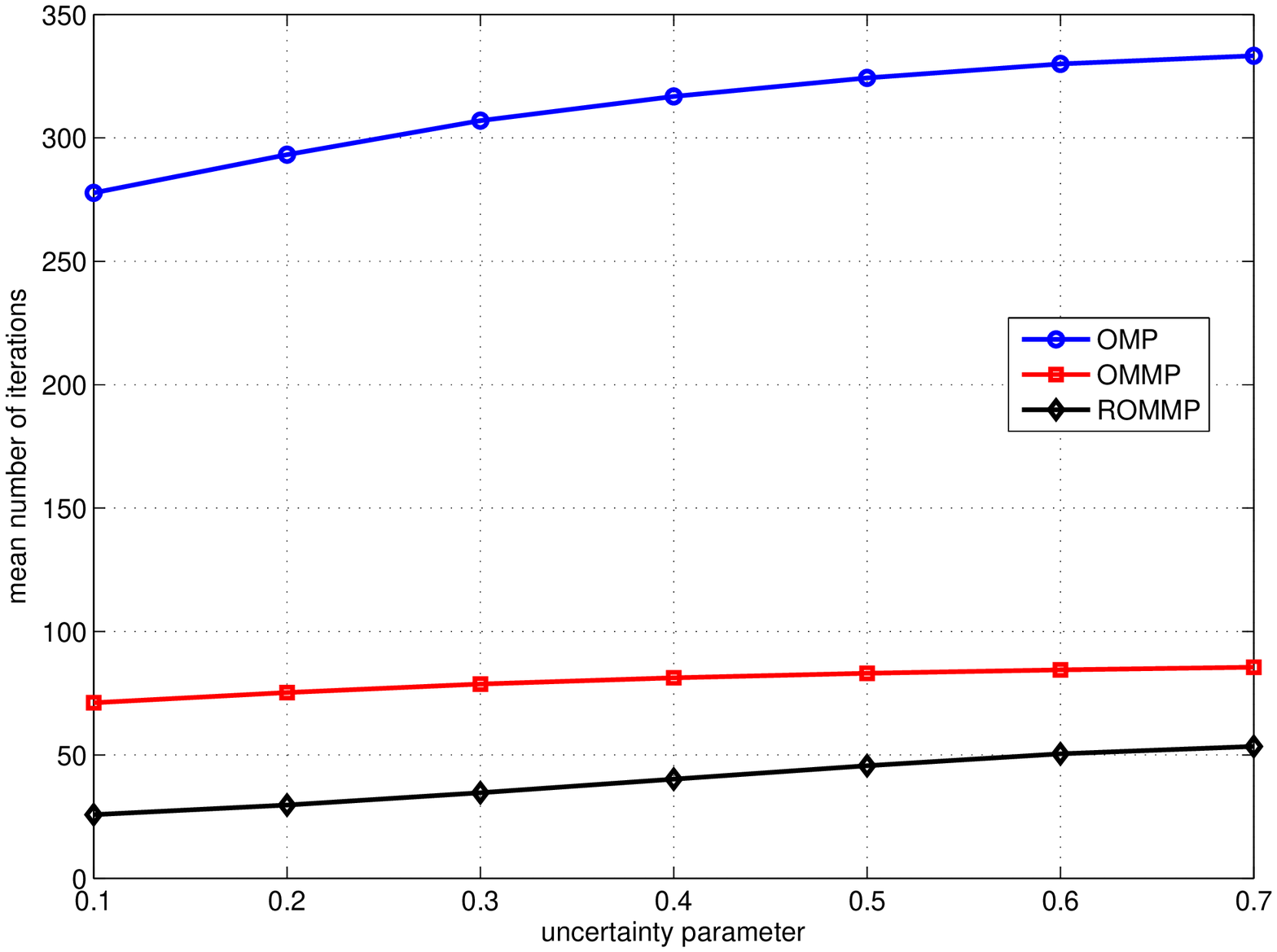}
                \caption{mean number of iterations vs uncertainty parameter}
                \label{fig:22-2}
        \end{subfigure}
\caption{Experiments of Group B: The mean number of iterations at various number of measurements and various uncertainty parameters of additive and multiplicative noise.}
\label{figure22}
\end{figure}

To compare the computational complexity, Fig.~\ref{figure22} shows the mean number of iterations at various percentages of additive and multiplicative noise and various number of compressive measurements. We can see that OMP needs the largest number of iterations. The proposed ROMMP needs less iterations than OMMP. Considering that the computational complexity of each iteration of OMP, OMMP, and ROMMP is almost the same, we can conclude that the ROMMP has the least computational complexity.

\section{Conclusion}
\label{sec6}

In this paper, we discuss the sampling and representation uncertainties in CS. A generalized sparse signal model considering both multiplicative noise and additive noise is given. Based on this model, a new optimization model for robust recovery of the generalized sparse signal is deduced by a stochastic analysis. Both convex relaxation and a greedy algorithm are used to solve the optimization model. Sufficient conditions for successful recovery are analyzed. Numerical experiments show that the proposed RL0 optimization based algorithms are in general superior to the previous ones.


%
%
%

 \newpage

\bibliographystyle{IEEEtran}

\end{document}